\documentclass[10pt, conference]{IEEEtran}
\IEEEoverridecommandlockouts
\usepackage{cite}
\usepackage{amsmath,amssymb,amsfonts}
\usepackage{algorithmic}
\usepackage{graphicx}
\usepackage{textcomp}
\usepackage{xcolor}
\def\BibTeX{{\rm B\kern-.05em{\sc i\kern-.025em b}\kern-.08em
    T\kern-.1667em\lower.7ex\hbox{E}\kern-.125emX}}
\usepackage{tabularx}

\usepackage{multirow}
\usepackage{adjustbox}

\usepackage{booktabs}
\usepackage{threeparttable}

\usepackage{siunitx}

\usepackage[%
 toc,              
 acronym,          
 section=chapter,  
 nonumberlist,     
]{glossaries}

\usepackage[hidelinks = true]{hyperref}

\newcommand{\PreserveBackslash}[1]{\let\temp=\\#1\let\\=\temp}
\newcolumntype{C}[1]{>{\PreserveBackslash\centering}p{#1}}

\DeclareSIPrefix\kilo{K}{3}

\usepackage{eso-pic}

\newcommand{\placetextbox}[3]{
  \setbox0=\hbox{#3}
  \AddToShipoutPictureFG*{
    \put(\LenToUnit{#1\paperwidth},\LenToUnit{#2\paperheight}){\vtop{{\null}\parbox[0pt][5pt][c]{\textwidth}{#3}}}%
  }%
}%

\glsdisablehyper

\makeglossaries

\glsaddall[types={main}]

\newacronym[plural=DNNs, firstplural={deep neural networks (DNNs)}]{dnn}{DNN}{deep neural network}
\newacronym[plural=RNNs, firstplural={recurrent neural networks (RNNs)}]{rnn}{RNN}{recurrent neural network}
\newacronym[plural=CNNs, firstplural={convolutional neural networks (CNNs)}]{cnn}{CNN}{convolutional neural network}
\newacronym{nlp}{NLP}{natural language processing}
\newacronym{dma}{DMA}{direct memory access}
\newacronym{mac}{MAC}{multiply-accumulate}
\newacronym{llc}{LLC}{Last-Level Cache}
\newacronym{spm}{SPM}{scratchpad memory}
\newacronym{simd}{SIMD}{single instruction, multiple data}
\newacronym{axi}{AXI}{Advanced eXtensible Interface}
\newacronym{soc}{SoC}{system-on-chip}
\newacronym{sram}{SRAM}{static random-access memory}
\newacronym{dram}{DRAM}{dynamic random-access memory}
\newacronym{hbm}{HBM}{High Bandwidth Memory}
\newacronym{cct}{CCT}{Compact Convolutional Transformer}
\newacronym[plural=PEs, firstplural={processing engines (PEs)}]{pe}{PE}{processing engine}
\newacronym[plural=FSMs, firstplural={finite state machines (FSMs)}]{fsm}{FSM}{finite state machine}
\newacronym[plural=LUTs, firstplural={lookup tables (LUTs)}]{lut}{LUT}{lookup table}
\newacronym[plural=CPUs, firstplural={central processing units (CPUs)}]{cpu}{CPU}{central processing unit}
\newacronym[]{fd_soi}{FD-SOI}{fully-depleted silicon-on-insulator}

\glsaddall[types={\acronymtype}]

\IEEEpubid{\makebox[\columnwidth]{ 979-8-3503-1175-4/23/\$31.00 $
\copyright$2023 IEEE \hfill{ \hspace{\columnsep}\makebox[\columnwidth]
{ }}}}

\begin{document}

\placetextbox{0.08}{0.05}{\textcolor{gray}{\footnotesize This paper has been accepted for publication by the International Symposium on Low Power Electronics and Design (ISLPED). \copyright 2023 IEEE. Personal use of this material is permitted.  Permission from IEEE must be obtained for all other uses, in any current or future media, including reprinting/republishing this material for advertising or promotional purposes, creating new collective works, for resale or redistribution to servers or lists, or reuse of any copyrighted component of this work in other works.}}

\title{ITA: An Energy-Efficient Attention and Softmax Accelerator for Quantized Transformers
}

\makeatletter
\newcommand{\linebreakand}{%
  \end{@IEEEauthorhalign}
  \hfill\mbox{}\par
  \mbox{}\hfill\begin{@IEEEauthorhalign}
}
\makeatother

\author{
\IEEEauthorblockN{Gamze Islamoglu\IEEEauthorrefmark{1}, Moritz Scherer\IEEEauthorrefmark{1}, Gianna Paulin\IEEEauthorrefmark{1}, Tim Fischer\IEEEauthorrefmark{1}, \\ Victor J.B. Jung\IEEEauthorrefmark{1}, Angelo Garofalo\IEEEauthorrefmark{1}\IEEEauthorrefmark{2}, Luca Benini\IEEEauthorrefmark{1}\IEEEauthorrefmark{2}} \IEEEauthorblockA{\IEEEauthorrefmark{1}ETH Zürich, Switzerland , \IEEEauthorrefmark{2}University of Bologna, Italy \\
\IEEEauthorrefmark{1}\{gislamoglu,scheremo,pauling,fischeti,jungvi,lbenini\}@iis.ee.ethz.ch, \IEEEauthorrefmark{2}angelo.garofalo@unibo.it}
}

\maketitle

\begin{abstract}

Transformer networks have emerged as the state-of-the-art approach for natural language processing tasks and are gaining popularity in other domains such as computer vision and audio processing. However, the efficient hardware acceleration of transformer models poses new challenges due to their high arithmetic intensities, large memory requirements, and complex dataflow dependencies. In this work, we propose ITA, a novel accelerator architecture for transformers and related models that targets efficient inference on embedded systems by exploiting 8-bit quantization and an innovative softmax implementation that operates exclusively on integer values. By computing on-the-fly in streaming mode, our softmax implementation minimizes data movement and energy consumption. ITA achieves competitive energy efficiency with respect to state-of-the-art transformer accelerators with \qty[detect-all=true]{16.9}{\tera{OPS}\per\watt}, while outperforming them in area efficiency with \qty[detect-all=true]{5.93}{\tera{OPS}\per\milli\square\meter} in \qty[detect-all=true]{22}{\nano\meter} fully-depleted silicon-on-insulator technology at \qty[detect-all=true]{0.8}{\volt}.

\end{abstract}


\vspace{-1em}

\section{Introduction}
The transformer is a deep learning architecture introduced in 2017 \cite{att_all}, which has revolutionized natural language processing tasks by achieving superior accuracy with respect to \glspl{rnn} at comparable compute and memory requirements. Recently, transformers have been adopted across multiple modalities, including text \cite{BERT, gpt2}, image \cite{vit}, audio \cite{audio}, and video \cite{video}. The ubiquity of the transformer model highlights its general-purpose capabilities \cite{ai_rep22} and stresses the need for efficient hardware acceleration. 

While most transformer models require gigabytes of memory for their parameters, and billions of operations for each inference, recent research has proven that smaller transformers have applications that suit low-power embedded systems \cite{mcu}. Besides architectural optimization, research into the compression of transformers has shown that 8-bit quantized models perform on par with their floating-point equivalents \cite{i-bert,Q8BERT}. 

A key component of transformers is the attention mechanism which generates a square matrix of order input length, resulting in a superlinear number of operations and memory size \cite{att_all}. This computation- and memory-intensive nature of the attention severely impacts the energy cost of deploying transformers on embedded systems, requiring specialized hardware to improve performance and energy efficiency. 

A peculiar challenge with transformers is the softmax operation which is applied over the rows of the attention matrix and becomes a bottleneck in low-precision architectures due to its nonlinear and non-element-wise nature. The nonlinearity of softmax restricts performing it on quantized values while the utilization of floating-point units incurs significant area and power costs. Furthermore, the non-element-wise nature of the softmax operation necessitates multiple passes through the attention matrix's row vectors, resulting in substantial data movement and power consumption within the system.

In this work, we present ITA, \underline{I}nteger \underline{T}ransformer \underline{A}ccelerator, an architecture targeting low-power embedded applications. To maximize ITA's energy efficiency, we focus on minimizing data movement throughout the execution cycle of the attention mechanism. In contrast to throughput-oriented accelerator designs, which typically employ systolic arrays, ITA implements its processing elements with wide dot-product units, allowing us to maximize the depth of adder trees, thereby further increasing efficiency.

To overcome the complex dataflow requirements of standard softmax, we present a novel approach that allows performing softmax on 8-bit integer quantized values directly in a streaming data fashion. Our approach also enables a weight stationary dataflow by decoupling denominator summation and division in softmax. The streaming softmax operation and weight stationary flow, in turn, minimize data movement in the system and power consumption of ITA. 

Our contributions can be summarized as follows:
\begin{itemize} 
    \item We present ITA, a hardware accelerator utilizing the parallelism of attention mechanism and 8-bit integer quantization to improve performance and energy efficiency. 
    To minimize data movement and power consumption, ITA adopts weight stationary dataflow over output stationary.
    \item We propose an energy- and area-efficient softmax implementation that fully operates in integer arithmetic with a footprint of only \qty[detect-all=true]{3.3}{\percent} over the total area of ITA and a mean absolute error of \qty[detect-all=true]{0.46}{\percent} compared to its floating-point implementation. The streaming operation further saves energy by reducing data movement.
    \item We evaluate our architecture in GlobalFoundaries' 22FDX \gls{fd_soi} technology and achieve an energy efficiency of \qty[detect-all=true]{16.9}{\tera{OPS}\per\watt} and area efficiency of \qty[detect-all=true]{5.93}{\tera{OPS}\per\milli\square\meter} at \qty[detect-all=true]{0.8}{\volt}, performing similarly to the state-of-the-art in energy efficiency, despite being implemented in a much less aggressive technology, and $2\times$ better in area efficiency.
\end{itemize}

\section{Preliminaries and Related Work}
In this section, we describe the operations in transformer-based networks, focusing on the softmax operation since it is a critical computation in transformers and creates a significant bottleneck in acceleration.

\subsection{Transformers}

A transformer network consists of multiple encoder and/or decoder stages, each containing an attention block, and a task-specific final layer. \autoref{fig:trafo} shows a transformer encoder and multi-head attention. In decoders, the inputs are slightly modified but the attention mechanism remains the same. 

\begin{figure}[]
  \centering
  \includegraphics[width=0.99\linewidth]{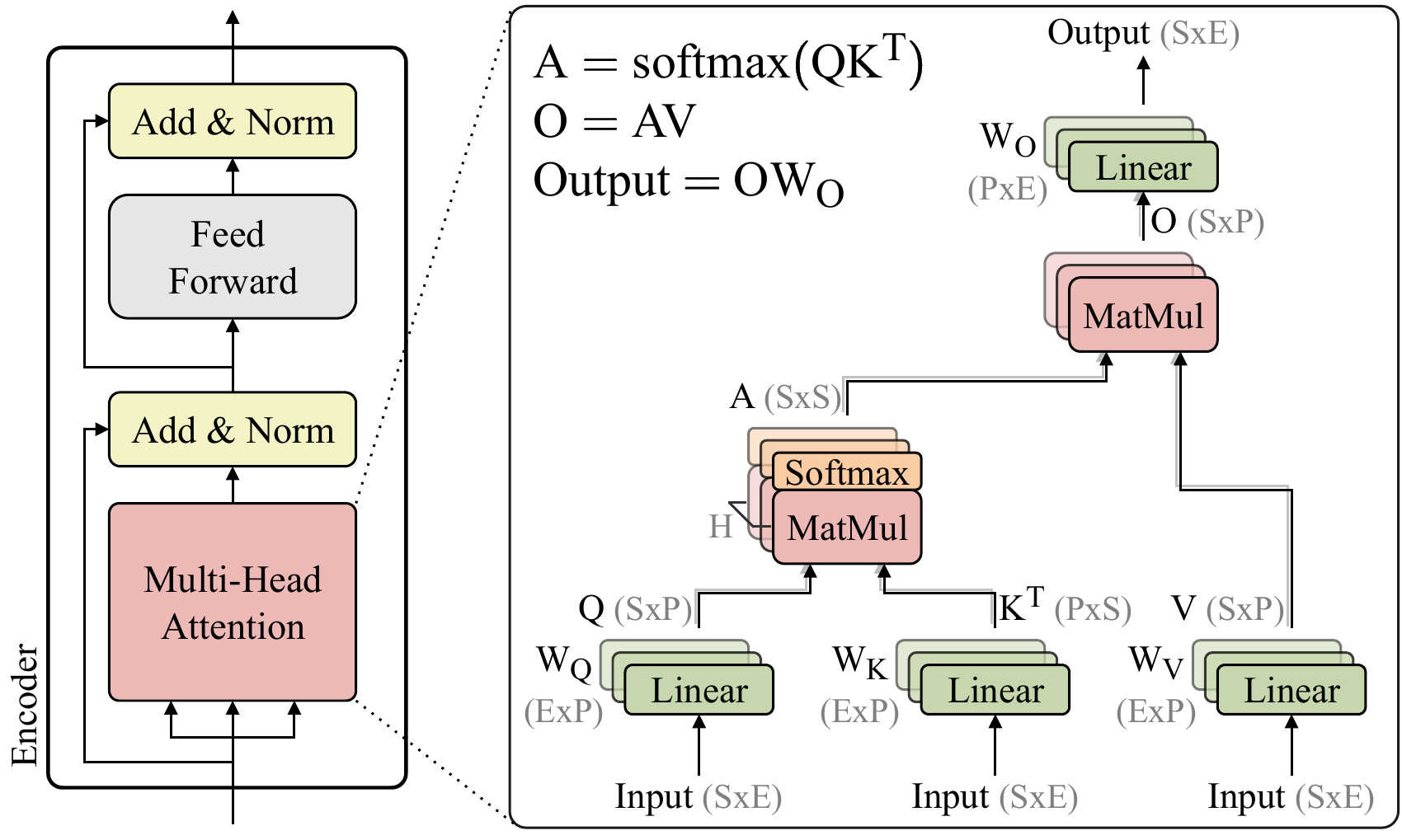}
  \caption{Transformer encoder and multi-head attention. S: sequence length, E: embedding size, P: projection space, H: number of heads.}
  \label{fig:trafo}
  \vspace{-1em}
\end{figure}

Multi-head attention is the main building block of transformers. In attention, three linear transformations are applied to inputs of size $S\times E$, where $S$ is the sequence length and $E$ is the embedding size, to generate Query ($Q$), \mbox{Key ($K$)}, and Value ($V$) matrices. $Q$, $K$, and $V$ are of size $S\times P$, where $P$ is the projection space. Then, matrix multiplication is performed between $Q$ and $K^T$, and softmax is applied to obtain probabilities. The resulting $S\times S$ attention matrix ($A$) can be considered as probabilities showing the relationship between Queries and Keys. $A$ is then multiplied by $V$, which weights the input tokens according to their relevance. By performing these operations in parallel with multiple sets of Query, Key, and Value matrices, multiple heads of attention are obtained. The outputs of these heads are then concatenated and linearly transformed to produce the final output of the attention, which has the same size as the input ($S\times E$).

\subsection{Softmax}
\label{softmax_bg}

Softmax is a key operation in transformers and encountered in every attention layer in the computation of matrix $A$. It is applied row-wise to the attention matrix to normalize it to probabilities. The softmax function is an $R^n \rightarrow R^n$ function, defined as follows for a vector $\boldsymbol{x}$ of length $n$:
\begin{equation}
    \text{softmax}(\boldsymbol{x})_i = \frac{e^{x_i-\max(\boldsymbol{x})}}{\sum_{j=1}^{n} e^{x_j-\max(\boldsymbol{x})}}
\end{equation}
and it produces a new vector of length $n$ whose elements sum to 1. Softmax presents two challenges: nonlinearity and non-element-wise operation. The nonlinearity means that we cannot perform softmax on the quantized values directly because $ \text{softmax}(\varepsilon x_q) \neq \varepsilon\cdot \text{softmax}(x_q)$ given the quantized value $x_q$ and scaling factor $\varepsilon$. In some accelerators, the input of softmax is first dequantized, softmax is calculated, and then output is quantized again \cite{Q8BERT, int8}. However, this approach is not hardware-friendly as it involves floating point units. \mbox{I-BERT~\cite{i-bert}} proposes a method to approximate softmax using second-order polynomials, eliminating the need for dequantization entirely. However, it operates at a higher precision of 32-bit, as opposed to the 8-bit quantization used in the rest of the network, and requires 32-bit multipliers and dividers.

Furthermore, softmax is not an element-wise operation and requires both a maximum search and a summation over a row of the attention matrix. This results in multiple passes over the row and multiple reads from memory, leading to high data movement and power consumption. Therefore, transformer accelerators usually compute the attention matrix row by row and accumulate the summation over the row. After completing one row, the division is performed to obtain probabilities \cite{besa, nvidia-4b}. However, this method is not feasible for weight stationary accelerators as the attention matrix is not produced row by row. ITA overcomes this issue and minimizes memory traffic by calculating a tight softmax approximation for 8-bit integers in three steps over multiple rows as explained in \autoref{sec:softmax}.

\subsection{Related Work}

Accelerating inference of transformer networks is an active area of research, with most accelerators focusing on the attention layer and using integer data formats like our approach. Some architectures exploit the sparsity of the attention matrix, such as OPTIMUS~\cite{optimus} which uses a sparse matrix format and redundant computation skipping in decoding. SpAtten~\cite{spatten} proposes token and head pruning and progressive quantization to reduce memory accesses and computations using a special engine to rank token and head importance scores. ELSA~\cite{elsa} utilizes an approximate self-attention algorithm to filter irrelevant query and key pairs and only performs exact computation for relevant pairs that are selected by hash and norm computation units. Similarly, Wang et al.~\cite{besa} propose a big-exact-small-approximate processing unit to save power and a bidirectional asymptotic speculation unit to skip redundant computations. However, the sparsity of transformers is limited to the attention matrix and depends on the network itself, and supporting sparsity in these accelerators comes with a cost in the area, such as additional top-k engine in SpAtten and hash and norm computation units in ELSA. Therefore, ITA does not utilize sparsity of attention to achieve higher area efficiency.

SpAtten and ELSA perform softmax in floating point by dequantizing before and quantizing after the softmax. However, this approach requires additional floating point units that are not utilized during the majority of computation, making it less preferable than integer equivalents due to larger area occupancy. Keller et al.~\cite{nvidia-4b} use the Softermax algorithm \cite{softermax}, which uses fixed-point arithmetic and replaces base $e$ with $2$ to simplify the hardware. In this paper, we present an alternative approach to compute softmax in integer with minimal area overhead in hardware, without approximating the softmax with base $2$. While Wang et al.~\cite{besa} and OPTIMUS~\cite{optimus} also compute softmax without conversion to floating point, they do not provide information about the implementation details and errors introduced by their softmax implementation.

\section{Architecture}
\begin{figure}[]
    \centering
    \includegraphics[width=0.99\columnwidth]{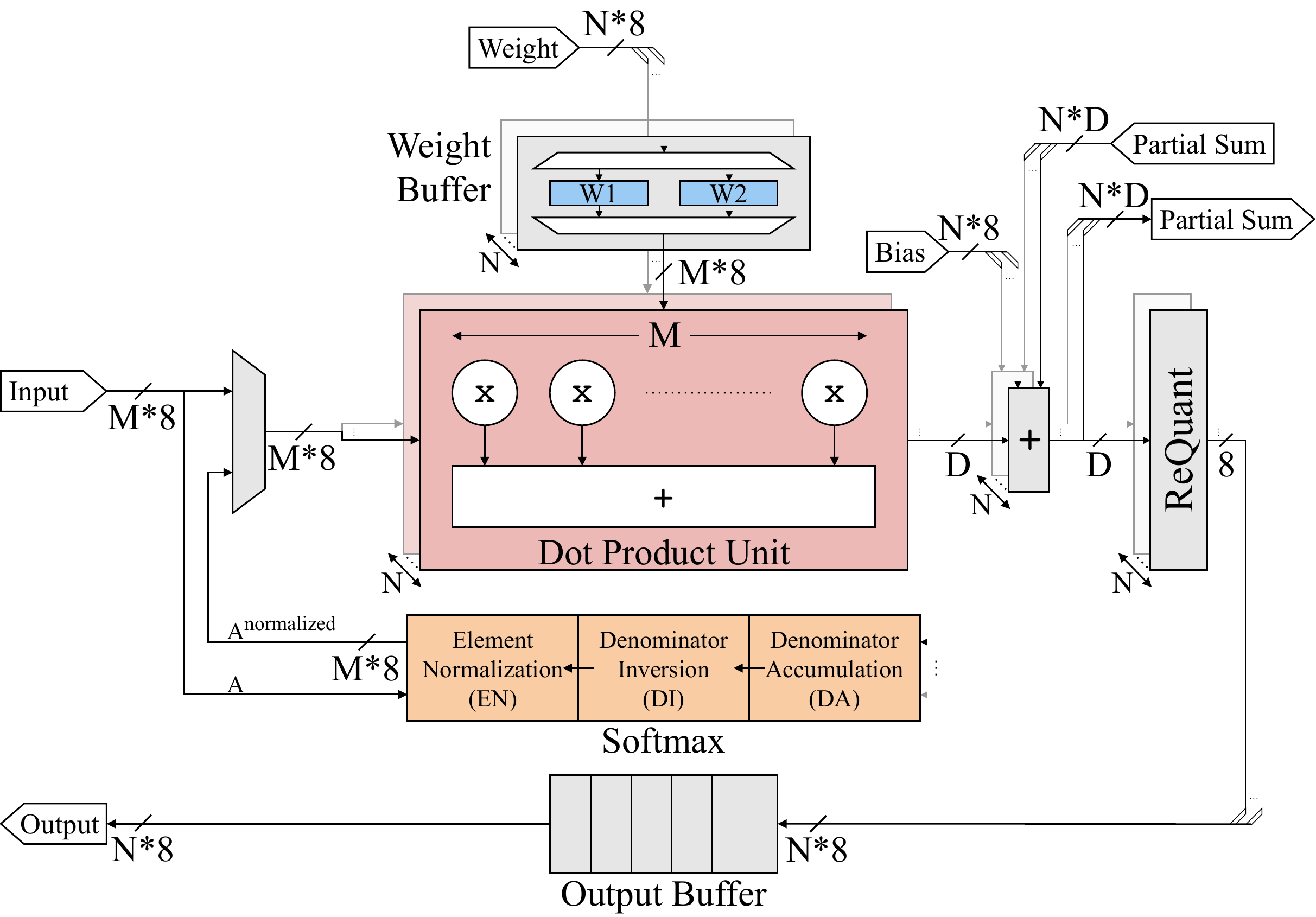}
    \caption{Architecture of ITA with 8-bit inputs and weights. The softmax block is detailed in \autoref{fig:softmax}.}
    \vspace{-1em}
    \label{fig:ita}
\end{figure}

The architecture of our transformer accelerator is shown in \autoref{fig:ita}, targeting 8-bit integer quantized matrices. The accelerator is parametric: it includes $N$ \glspl{pe}, each computing the dot product between two vectors of $M$ elements, and works on tiles of size $M\times M$. Each \gls{pe} uses 8-bit weights and activations, producing dot product results with higher precision of $D$-bit. $N$, $M$, and $D$ are configured at design time. The adders after \glspl{pe} accumulate partial sums. Once outputs are fully accumulated, 8-bit biases are added to outputs, which are then converted back to 8-bit format by requantization modules (ReQuant in \autoref{fig:ita}). 

The softmax module computes the softmax of the attention matrix $A$ and works in two passes. In the first pass, when it takes the elements of $A$ from the matrix multiplication $ Q \times K^T$, it finds the maximum and accumulates the denominator of softmax. In the second pass, when the attention matrix is supplied as input for the $A \times V$ computation, the softmax module normalizes them to probabilities before entering \glspl{pe}.
To achieve high throughput and low power consumption, we propose a novel and hardware-friendly softmax implementation, which is detailed in \autoref{sec:softmax}. The explained clipping operation is performed by the requantization module and the clipping threshold is obtained from quantization-aware training that incorporates our softmax implementation.

Finally, the output FIFO buffers the results temporarily to prevent stalling the accelerator in case the output cannot be written to the memory immediately. 

ITA follows a weight stationary approach to reduce the bandwidth and energy requirements.  Weights are reused $M$ times and stored in a double-buffered weight buffer, where $W1$ and $W2$ have a capacity of $M$ bytes. Double buffering allows the accelerator to fetch weights for the next computation while simultaneously performing the current computation. This reduces the bandwidth requirement for the weight interface from $NM$ to $N$ bytes per cycle. 
While the weight stationary approach of ITA only requires a bandwidth of $8(M+3N)+2ND$ bits per cycle ($M$ bytes read for input, $N$ bytes read for weight and bias, $N$ bytes write for output, and $ND$ bits read and $ND$ bits write for partial sums), output stationary approaches typically require substantially more at $8(NM+3N)+2ND$ bits per cycle ($NM$ bytes read for weight, $N$ bytes read for input and bias, $N$ bytes write for output, and $ND$ bits read and $ND$ bits write for partial sums). As the number of processing elements in the accelerator increases, ITA can sustain higher utilization compared to an output stationary flow, with fewer data movements leading to lower power consumption. However, the downside of this approach is the size of the weight buffer ($2NM$ bytes). An output stationary accelerator can double-buffer inputs with a buffer size of $2M$ bytes without buffering weights since they are updated every cycle. We prefer the former because memory bandwidth is often the bottleneck, especially for accelerators, since only a small portion of network parameters can be stored locally and they have to access higher levels of memory continuously.
Another difficulty of weight stationary flow is the row dependency of softmax, as explained in \autoref{softmax_bg}. In \autoref{sec:softmax}, we discuss our proposed method to handle this dependency.

\begin{figure}[b]
    \centering
    \vspace{-1em}
    \includegraphics[width=0.99\columnwidth]{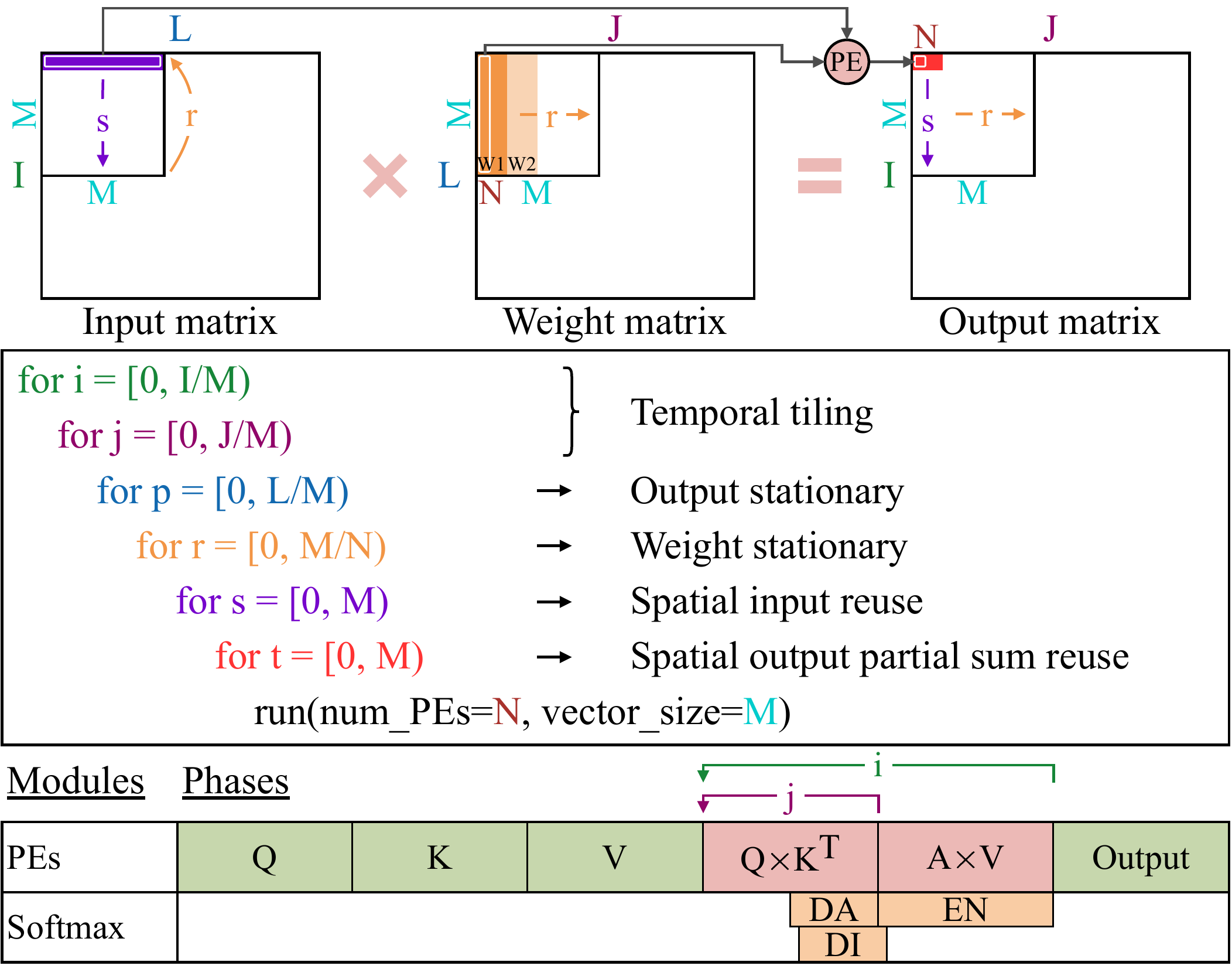}
    \caption{Workload mapping and computation phases.}
    \label{fig:tiling}
    \vspace{-0em}
\end{figure}

The workload mapping and schedule of ITA are summarized in \autoref{fig:tiling}. The accelerator operates on tiles of size $M\times M$ and iterates over dimension $L$ to achieve output stationarity in the outer loop. Within each tile, ITA employs a weight stationary regime and shares inputs among $N$ \glspl{pe}, achieving spatial input reuse. Each \gls{pe} operates on vectors of size $M$ in the innermost loop and computes the dot product of input and weight vectors. If $M$ is not an integer multiple of matrix dimensions, inputs/weights are padded with zeros. 

\begin{figure*}[t]
    \centering
    \includegraphics[width=\textwidth]{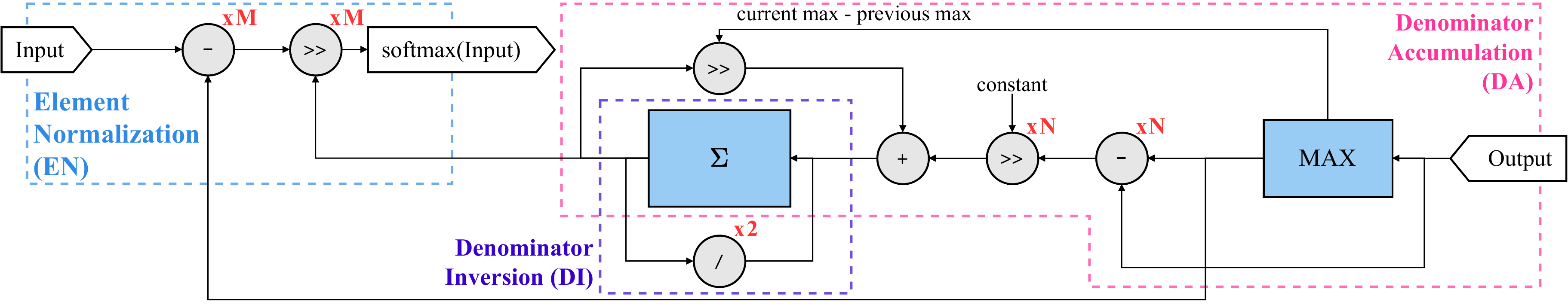}
    \caption{Softmax implementation. Buffers are shown in blue.}
    \label{fig:softmax}
    \vspace{-1em}
\end{figure*}

ITA computes linear layers sequentially but fuses $Q\times K^T$ and $A\times V$ in iterations of $i$. In each final iteration of a $Q\times K^T$ tile, the softmax module accumulates denominators partially (\textit{DA}). Once a row of $M\times J$ section of the attention matrix is completed, the softmax module inverts the denominator of the row (\textit{DI}) and stores the inverted denominator. Then, $M$ rows of $A\times V$ are computed while normalizing elements of $A$ in the softmax module (\textit{EN}). At the start of the next iteration of $i$, the softmax module is reset and the same steps are repeated until all iterations are completed.

\section{Softmax}
\label{sec:softmax}

We propose a novel hardware-friendly implementation of softmax, shown in \autoref{fig:softmax}, with the following features:

\begin{itemize}
    \item The softmax is computed on the quantized values directly.
    \item To prevent underflow, both the nominator and denominator are scaled with an integer value. Therefore, the accumulation and inversion of the denominator are performed in 15-bit and 16-bit integer formats, respectively.
    \item We add minimal memory overhead to store the maximum and sum values. Both maximum and sum buffers contain $M$ elements, equal to the number of rows of a tile.
    \item We do not use any exponentiation modules and multipliers which are costly in terms of area and power.
    \item Softmax is computed on-the-fly and does not add any latency to the computation as shown in \autoref{fig:tiling}.
    \item By computing softmax on streaming data, we avoid fetching the same vector multiple times, reducing the data movement and power consumption. 
\end{itemize}

Our main observation is that above a certain value of the scaling factor, softmax quantizes to zero for all inputs except for the maximum of the input. This means that the range of the inputs can be clipped to the inputs that will end up with a softmax greater than 0, and the scaling factor can be tuned accordingly in training time as shown in \autoref{fig:softmax_clip}. Secondly, we can hide the factor $\log_2 e$ in the scaling factor $\varepsilon$ and change the base to $2$ to simplify the hardware, as follows: 
\begin{equation}
    e^{x} = e^{\varepsilon x_q} = (2^{\log_2 e})^{\varepsilon x_q} = 2^{((\log_2 e) \varepsilon ) x_q} = 2^{\varepsilon^\prime x_q}
\end{equation}
where $x_q$ is the quantized value ($x=\varepsilon x_q$) and $\varepsilon^\prime = (\log_2 e) \varepsilon$.

\begin{figure}[b]
    \centering
    \vspace{-1em}
    \includegraphics[width=0.95\linewidth]{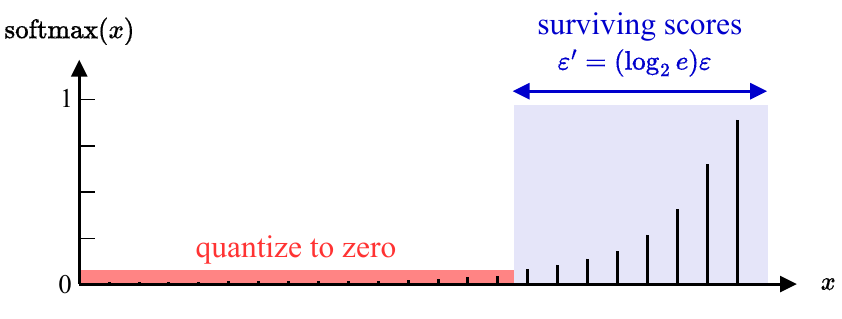}
    \caption{Effect of softmax and quantization on attention probabilities.}
    \label{fig:softmax_clip}
    \vspace{0em}
\end{figure}

The maximum meaningful scaling factor, computed based on the range of inputs with non-zero quantized softmax, is $\varepsilon = B/(2^B \log_2 e)$, where $B$ is the number of bits used in quantized representation (equals 8 in our case). Using this scaling factor, $\varepsilon^\prime$ becomes:
\begin{equation}
    \varepsilon^\prime = (\log_2 e) \varepsilon = (\log_2 e) B/(2^B \log_2 e) = B/2^B
\end{equation}
and softmax can be written as follows:
\begin{equation} \label{eq:softmax}
    \begin{split}
    \text{softmax}(\boldsymbol{x})_i & = \frac{2^{\varepsilon^\prime  (x_{qi} - \max(\boldsymbol{x_q}))}}{\sum_{j=1}^{n} 2^{\varepsilon^\prime  (x_{qj} - \max(\boldsymbol{x_q}))}} \\
    & = \frac{2^{\frac{B}{2^B}  (x_{qi} - \max(\boldsymbol{x_q}))}}{\sum_{j=1}^{n} 2^{\frac{B}{2^B}  (x_{qj} - \max(\boldsymbol{x_q}))}} \\
    & = \frac{2^{(x_{qi} - \max(\boldsymbol{x_q})) \gg (B-\log_2 B)}}{\sum_{j=1}^{n} 2^{(x_{qj} - \max(\boldsymbol{x_q})) \gg (B-\log_2 B)}} \\
    \end{split}
\end{equation}

Using the above formula, the softmax module is implemented as depicted in \autoref{fig:softmax} and softmax is computed in three steps as shown in \autoref{fig:tiling}. 
In \textit{Denominator Accumulation (DA)}, we find the maximum of the first computed part of a row and store it in the MAX buffer. Then, we subtract the maximum from all the elements, accumulate the sum, and store it in the \textit{$\Sigma$} buffer. When we get the next parts of the row, we compare the previous maximum that is stored in the MAX buffer with the current maximum. If the current maximum is greater, we update the maximum. The difference between the two maximums is used to update the accumulated sum in \textit{$\Sigma$} and added up with the accumulation over the current part of the row. These operations are repeated over $M$ rows of the attention matrix and the maximum and accumulated sum are stored in the respective buffers for each row. Once the denominator of the softmax is accumulated for a row in \textit{DA}, the inverse of the denominator is computed in \textit{Denominator Inversion (DI)} using serial dividers and stored in the \textit{$\Sigma$} buffer.
Since \textit{DI} is overlapped with \textit{DA}, we have plenty of time to compute the inverse of the denominator. Therefore, only two serial dividers suffice to compute the inverse without causing any stalls.
After obtaining the inverse of the denominator ($\Sigma_{inverse}$), we compute the softmax by shifting it as follows in \textit{Element Normalization (EN)}:
\begin{multline}
    \text{softmax}(\boldsymbol{x})_i=\Sigma_{inverse} \gg \\
    \left ((\max(\boldsymbol{x_q})- x_{qi}) \gg (B-\log_2 B)\right )
\end{multline}
As $B=8$ is a constant in the architecture, a programmable shifter is not required for shifting by $B-\log_2 B$. Here, $B-\log_2 B$ evaluates to $8-\log_2 8=5$, and we can simply take the most significant 3 bits of $(\max(\boldsymbol{x_q})- x_{qi})$ to  perform the shift.

\vspace{-0.5em}

\section{Evaluation}
\subsection{Physical Implementation and Measurements}

We evaluate ITA with 16 processing engines consisting of 64 \gls{mac} units ($N=16$ and $M=64$) and $D$ is selected 24-bit to allow up to 256-element dot products, enough for the targeted compact models \cite{vit}.  The memory buffers for weights and for storing the maximum and sum values in the softmax module are made of latch-based memories and clock-gated.

ITA is implemented in GlobalFoundries' 22FDX FD-SOI technology and targets an operating frequency of \qty[detect-all=true]{500}{\mega\hertz} in worst-case conditions (SS/\qty[detect-all=true]{0.72}{\volt}/\qty[detect-all=true]{125}{\celsius}). Synopsys' Fusion Compiler 2022.03 is employed for both synthesis and implementation of the accelerator. The power consumption of ITA is estimated using Synopsys' PrimeTime 2022.03, which takes into account the switching activities obtained from a post-layout gate-level simulation using a synthetic benchmark at the operating frequency of \qty[detect-all=true]{500}{\mega\hertz}. The power consumption is estimated under typical conditions (TT/\qty[detect-all=true]{0.80}{\volt}/\qty[detect-all=true]{25}{\celsius}).

\subsection{Experimental Results}

The total area occupied by ITA is \qty[detect-all=true]{0.173}{\milli\square\meter}. The area breakdown of ITA is presented in \autoref{fig:power_area}. The \glspl{pe} take \qty[detect-all=true]{58.1}{\percent} of the total area, while the weight buffer occupies \qty[detect-all=true]{19.6}{\percent}. Others include the remaining components of ITA's datapath (\qty[detect-all=true]{6.3}{\percent}), control circuitry (\qty[detect-all=true]{2.3}{\percent}), and output buffer (\qty[detect-all=true]{1.1}{\percent}). The hardware-friendly softmax solution implemented in this work proves to be very area efficient, with only \qty[detect-all=true]{3.3}{\percent} area contribution, corresponding to \qty[detect-all=true]{28.7}{\kilo{GE}}.

The entire accelerator consumes a total power of \qty[detect-all=true]{60.5}{\milli\watt} over the execution of attention. \autoref{fig:power_area} shows the power breakdown of ITA. The majority of power is consumed in \glspl{pe} with \qty[detect-all=true]{59.5}{\percent}. Clock tree and I/O registers (\qty[detect-all=true]{22.9}{\percent}) also lead to significant power consumption due to their high toggling rate. Others consist of remaining datapath elements of ITA (\qty[detect-all=true]{6.7}{\percent}), weight buffer (\qty[detect-all=true]{1.7}{\percent}) and output buffer (\qty[detect-all=true]{0.7}{\percent}). The softmax module only consumes \qty[detect-all=true]{1.4}{\percent} of the power. Although the weight buffer of ITA takes a significant portion of the area, its power consumption is less than \qty[detect-all=true]{2}{\percent} due to clock-gating. 

\begin{figure}[]
    \centering
    \includegraphics[width=0.85\columnwidth]{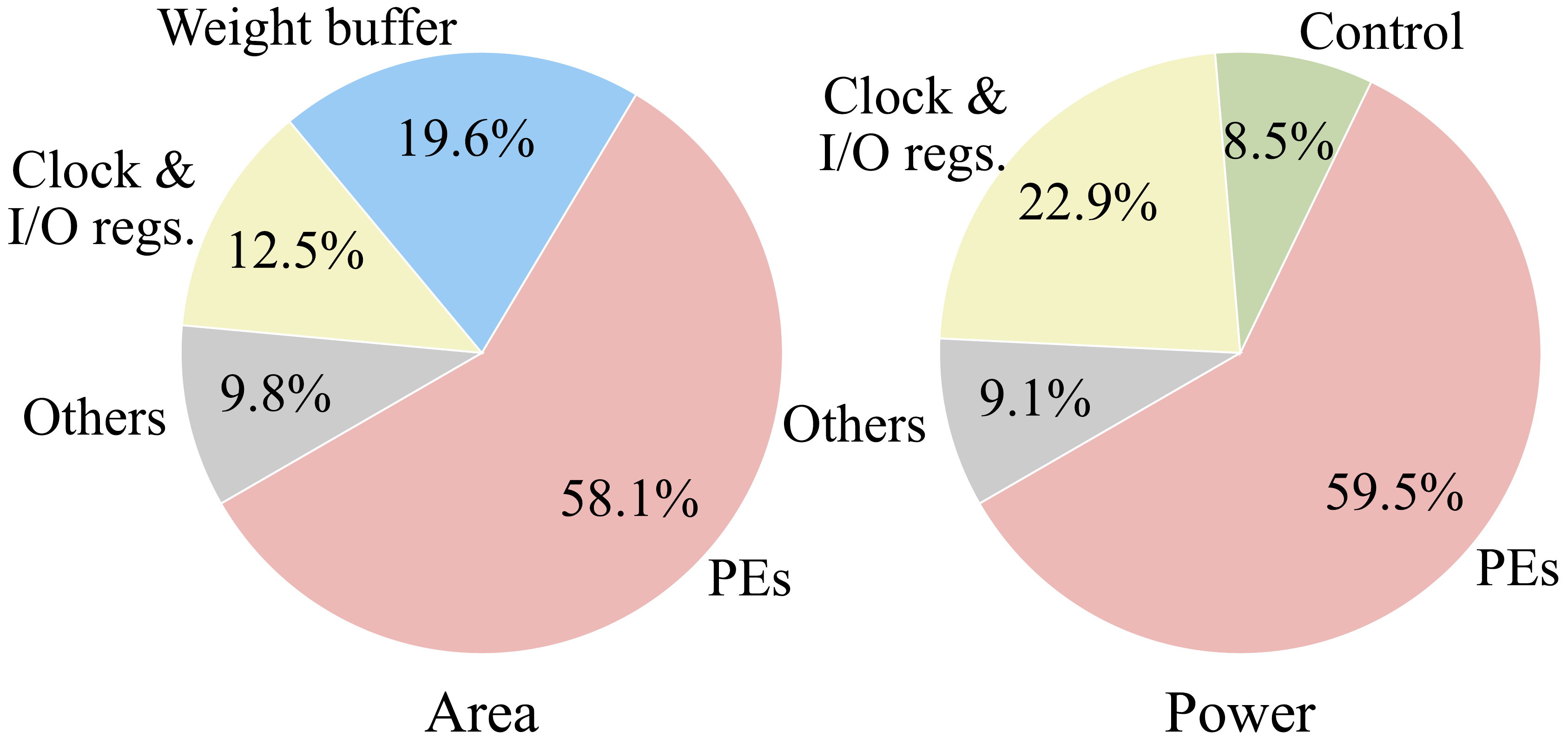}
    \caption{Area and power breakdown of ITA.}
    \label{fig:power_area}
    \vspace{-1.4em}
\end{figure}

\subsection{Softmax}

To assess the accuracy of our softmax implementation, we compare the Mean Absolute Error (MAE) of our implementation with the 32-bit integer-only softmax from I-BERT \cite{i-bert}. We use the activation of the Compact Transformer \cite{cct} as input in order to simulate the data distribution of a real transformer. Our implementation achieves an MAE of $4.6\mathrm{e}{-3}$, meaning that the average distance to the floating point value is \qty[detect-all=true]{0.46}{\percent}. The MAE of I-BERT's softmax is \qty[detect-all=true]{0.35}{\percent}, the slightly lower MAE is explained by the difference in input precision (32-bit for I-BERT vs 8-bit for ours). Compared to the I-BERT implementation which uses 32-bit multipliers and dividers, our approach operates at a lower precision and features a much simpler datapath, resulting in better latency and energy consumption.

\subsection{Performance Evaluation}

We compare ITA with a software baseline executed on MemPool, consisting of 256 32-bit RISC-V cores with \gls{simd} support \cite{mempool}. We use a highly optimized kernel for matrix multiplications and the I-BERT algorithm for softmax. Compared to MemPool, ITA achieves $6\times$ speedup and $45\times$ energy efficiency in attention computation.

\subsection{Comparison to State-of-the-Art} \label{sec:comparison}

\begin{table*}[t]
    \centering
    \caption{Comparison of the proposed architecture to state-of-the-art transformer accelerators.}
    \label{tab:comparison}
    \begin{adjustbox}{max width=\textwidth}
    \begin{threeparttable}
    \begin{tabular}{lcccccccc}
    \toprule
    & \multirow{2}{*}{ OPTIMUS \cite{optimus}} & \multirow{2}{*}{SpAtten \cite{spatten}} & \multirow{2}{*}{ELSA \cite{elsa}} & \multirow{2}{*}{Wang et al. \cite{besa}} & \multicolumn{2}{c}{Keller et al. \cite{nvidia-4b}} & \multicolumn{2}{c}{\textbf{This work}}
    \\\cmidrule(lr){6-7} \cmidrule(lr){8-9}
    & & & & & INT4 & INT8 & ITA & ITA System
    \\ \midrule
    \textbf{Technology} [\unit[detect-all=true]{\nano\metre}]
    & 28 & 40 & 40 & 28 & 5 & 5 & 22 & 22
    \\  
    \textbf{Area} [\unit[detect-all=true]{\milli\metre^2}]
    & 5.2 & 18.71 & 1.26/2.15\tnote{1} & 6.82 & 0.153 & 0.153 & 0.173 & 0.407
    \\  
    \textbf{Voltage} [\unit[detect-all=true]{\volt}]
    & 1 & 1.1 & 1.1 & 0.56-1.1 & 0.46-1.05 & 0.46-1.05 & 0.8 & 0.8
    \\ 
    \textbf{Frequency} [\unit[detect-all=true]{\mega\hertz}]
    & 200 & 1000 & 1000 & 50-510 & 152-1760 & 152-1760 & 500 & 500
    \\ 
    \textbf{Data formats}
    & INT16 & INT8-16/FP & INT8/FP16 & INT12 & INT4 & INT8 & INT8 & INT8 
    \\ 
    \textbf{Number of MAC units} 
    & 1024 & 1024 & 528 & 512 & 1024 & 512 & 1024 & 1024
    \\
    \textbf{On-chip memory} [\unit[detect-all=true]{\kilo\byte}]
    & 1420 & 392 & 4.5/148\tnote{1} & 336 & 141 & 141 & 2.24 & 67.8
    \\ 
    \textbf{Power} [\unit[detect-all=true]{\milli\watt}]
    & 731.8 & 2600 & 969.4/1494.2\tnote{1} & 12.06-272.8 & - & - & 60.5 & 121
    \\ 
    \textbf{Throughput} [\unit[detect-all=true]{\tera{OPS}}]
    & 0.5 & 1.61 & 1.09 & 0.52-4.07\tnote{2} & 3.6\tnote{4} & 1.8\tnote{4} & 1.02 & 1.02
    \\
    \textbf{Energy efficiency} [\unit[detect-all=true]{\tera{OPS}\per\watt}]
    & 0.68 & 0.62 & 1.12/0.73\tnote{1} & 1.91-27.56\tnote{3} & \textbf{95.6}\tnote{5} & \textbf{39.1}\tnote{5} & 16.9 & 8.46
    \\ 
    \textbf{Area efficiency} [\unit[detect-all=true]{\tera{OPS}\per\milli\metre^2}]
    & 0.096 & 0.086 & 0.865/0.507\tnote{1} & 0.0765-0.597\tnote{3} & \textbf{23.3}\tnote{4} & \textbf{11.7}\tnote{4} & 5.93 & 2.52
    \\
    \textbf{Area efficiency\tnote{*}} [\unit[detect-all=true]{\tera{OPS}\per\mega{GE}}]
    & 0.0310 & 0.0566 & 0.569/0.333\tnote{1} & 0.0247-0.192\tnote{3} & 0.242\tnote{4} & 0.121\tnote{4} & \textbf{1.18}	& \textbf{0.500}	
    \\ \bottomrule           
    \end{tabular}
    \begin{tablenotes}
        \item[*] Gate-equivalents (GE) of other technologies are scaled based on the GE of \qty[detect-all=true]{22}{\nano\metre} technology.
        \item[1] with external memory modules (on-chip SRAM).
        \item[2] measured at \qty[detect-all=true]{1.1}{\volt}, \qty[detect-all=true]{510}{\mega\hertz}. The highest throughput is measured for $A\times V$ assuming \qty[detect-all=true]{90}{\percent} near-zero and zero probabilities.
        \item[3] the lowest value measured at \qty[detect-all=true]{1.1}{\volt}, \qty[detect-all=true]{510}{\mega\hertz} for linear layers that compute $Q, K, V$ and the highest value measured at \qty[detect-all=true]{0.56}{\volt}, \qty[detect-all=true]{50}{\mega\hertz} for $A\times V$ assuming \qty[detect-all=true]{90}{\percent} near-zero and zero probabilities.
        \item[4] at \qty[detect-all=true]{1.05}{\volt}.
        \item[5] at \qty[detect-all=true]{0.46}{\volt}.
      \end{tablenotes}
    \end{threeparttable}
    \end{adjustbox}
    \vspace{-1.5em}
\end{table*}

We present a comparison of ITA to state-of-the-art transformer accelerators in \autoref{tab:comparison}. To have a fair comparison, we evaluate ITA as a standalone accelerator and integrate it into a system with \qty[detect-all=true]{64}{\kilo{i}\byte} \gls{sram}. The latter we call ITA System.

ITA achieves an energy efficiency of \qty[detect-all=true]{16.9}{\tera{OPS}\per\watt} standalone, and \qty[detect-all=true]{8.46}{\tera{OPS}\per\watt} integrated into the system, which is superior to all other accelerators except for Keller et al. \cite{nvidia-4b}. If we hypothetically scale down the voltage to 0.46V, using $V_{dd}^2$ scaling, ITA would be $1.3\times$ more efficient, and the system would be only $1.5\times$ less efficient than \cite{nvidia-4b}, despite being implemented in much less advanced technology. Wang et al. \cite{besa} report higher efficiency in 12-bit, but only at lower voltage and with the assumption of \qty[detect-all=true]{90}{\percent} sparsity. Furthermore, this sparsity exploitation reduces the area efficiency, which is much lower than ITA not only because of higher precision but also because of additional speculation and out-of-order execution logic.

ITA outperforms all other accelerators in terms of area efficiency, except for Keller et al.'s accelerator \cite{nvidia-4b}, which uses a \qty[detect-all=true]{5}{\nano\metre} technology. To provide a technology-independent metric, we present the area efficiency in terms of gate-equivalent (GE) as well and ITA surpasses all accelerators both as a standalone accelerator and at the system level.

\section{Conclusion}
We presented ITA, a hardware accelerator for quantized transformers that exploits parallelism of attention mechanism and 8-bit integer quantization to achieve efficient inference on embedded systems. Our architecture features a novel and hardware-friendly softmax implementation that operates directly on quantized values and facilitates weight stationary dataflow, reducing power consumption. ITA is evaluated on an advanced \qty[detect-all=true]{22}{\nano\meter} technology, achieving energy efficiency of \qty[detect-all=true]{16.9}{\tera{OPS}\per\watt} and area efficiency of \qty[detect-all=true]{5.93}{\tera{OPS}\per\milli\square\meter}. 

\section*{Acknowledgment}

This work is supported in part by the NeuroSoC project funded under Horizon Europe Grant Agreement n° 101070634.

\bibliographystyle{IEEEtran}
\bibliography{IEEEabrv,main}

\end{document}